\documentclass[12pt]{article}
\usepackage[pdftex]{graphicx}
\usepackage{color}
\newcommand{\comment}[1]{}
\newcommand{\commentu}[1]{}
\begin{document}
\newcommand{\be}{\begin{equation}}
\newcommand{\ee}{\end{equation}}
\parskip 10pt
\parindent 20pt

\title{Baby Morse Theory in Data Analysis}
\author{Caren Marzban$^{1,2}$\thanks{Corresponding Author:
marzban@stat.washington.edu} and Ulvi Yurtsever$^{3,4}$  \\
$^1$ Applied Physics Laboratory, and $^2$ Department of Statistics,\\
 University of Washington, Seattle, WA 98195,\\
$^3$ MathSense Analytics, 1273 Sunny Oaks Circle, Altadena, CA 91001  \\
$^4$ Hearne Institute for Theoretical Physics, \\
Louisiana State University, Baton Rouge, LA 70803 
}
\date{ }
\maketitle

\newpage
\begin{abstract}

A methodology is proposed for inferring the topology underlying
point cloud data. The approach employs basic elements of Morse
Theory, and is capable of producing not only a point estimate
of various topological quantities (e.g., genus), but it can also
assesses their sampling uncertainty in a probabilistic fashion. 
Several examples of point cloud data in three 
dimensions are utilized to demonstrate how the method yields
interval estimates for the topology of the data as a 2-dimensional surface
embedded in $R^3$. \\

\parindent 0pt
Key Words: Topology, Morse theory, persistence, inference, signal detection.

\parindent 20pt

\end{abstract}

\newpage
\section{Introduction}

There are many sources of high-dimensional data that are inherently structured 
but where the structure is difficult to conceptualize. The motivation to organize, 
associate, and connect multi-dimensional data in order to qualitatively 
understand its global content has recently led to the development of new tools 
inspired by topological methods of mathematics~\cite{carlsson,niyogi,pasc,zomo}.
The applications of topological 
data analysis methods include dimensionality reduction~\cite{leever}, 
computer vision~\cite{pasc}, and shape discovery~\cite{adan}. 
In most of these applications, the data is {\it point cloud data}, i.e., 
the coordinates of points in some space. Such data arise naturally 
in LIDAR (Light Detection and Ranging)~\cite{grejner}, 
image reconstruction~\cite{haji}, and in the geosciences~\cite{waw}.
In addition, point-cloud data in multidimensional Euclidean space can arise
from nonlinear transforms of other kinds of data such as time series~\cite{gilmore}.

Consider, for example, a cloud of points in 3-dimensional Euclidean space.
The cloud of points may be confined mostly to the surface of a
2-dimensional sphere; or to the surfaces of multiple disconnected spheres.
The number of such spheres is an example of a topological quantity, in
contrast to the specific shape of the spheres (e.g. round vs. squashed)
which is a geometrical quantity. Another example of a topological
quantity is the number of handles; a sphere has none, but the surface
of a doughnut has one. A sphere and a doughnut are
topologically distinct surfaces in that one cannot be transformed to
the other without cutting and gluing operations. The number of handles,
known as {\it genus}, is important because it turns out any 2-dimensional
compact surface can be constructed by gluing handles onto a sphere~\cite{lee}.
Said differently, the genus is a defining characteristic of the
topology of a 2-dimensional compact surface. As a final example, 
compare the surface of a ball with that of a coffee mug;
the former has genus 0, while the latter has genus 1. This type of
topological information can be useful in correctly identifying
underlying structures in point cloud data.

Whereas the human eye is capable of inferring such structures, one 
often requires a method for performing that task objectively. For
instance, the high dimensionality of the data may not allow visualization
in 3 dimensions. Even in 3 dimensions, it may be that the topological 
structure must be inferred in a streaming environment, where a human 
operator cannot visually inspect every situation one at a time. Finally,
there may be situations wherein the existence of an underlying structure
is not unambiguously evident even to a human expert. In such a situation, 
an algorithm capable of assigning probabilities to the various topological 
structures can be useful for decision making~\cite{katz}.

Inferring the various disconnected components of any structure can be done 
via a class of statistical methods generally known as cluster 
analysis~\cite{everitt}.
Some cluster analysis methods are also naturally capable of assigning
probabilities to the different number of components/clusters. However,
such methods are incapable of inferring higher-order topological structures.
For instance, no clustering algorithm can identify the number of handles 
(i.e. genus) of a 2-dimensional surface underlying point cloud data. It is that
task which is addressed in this paper. The method also produces
probabilities for the various possible genus values.

Two main methodologies of topological data analysis have been discussed
so far in the literature: One is based on the idea of persistence, and the other
on discretized approaches
to Morse Theory. This paper discusses a new approach to discrete Morse
Theory, illustrated by analyzing simple examples of point-cloud data in
three-dimensional Euclidean space. The alternative methodology,
based on persistence, utilizes the idea of capturing topological features in data
by analyzing continuous structures which are associated to data points as a function of a varying
scale parameter that measures, roughly, how coarsely the data points are
assumed to sample an underlying topological manifold.

To briefly illustrate the idea of persistence, assume a given data set $D$ 
consisting of a sampling of a smooth submanifold $M \subset R^n$
of $n$-dimensional Euclidean space. Evolution has
moulded the human perception system with the
ability to reconstruct geometric information
from two-dimensional projections; but this capability is only useful
in dimension $n=3$. For submanifolds embedded
in higher-dimensional Euclidean space $R^n$ ($n \geq 4$),
global features of the submanifold $M \subset R^n$ cannot be
read out from visual inspection of two-dimensional projections.
One topological invariant that is immensely useful in ascertaining
the ``global configuration" of the surface $M$, and therefore the true 
global nature of
the ``model" $M$ of the given data $D$, is the sequence of homology
groups $H_k (M)$, further described below. As mentioned previously,
some statistical methodologies like clustering
can be viewed as methods for the extraction of homological information.
Persistence is a general method to extract homology information
from a given data set. The natural question persistence attempts to
answer is: how can one compute $H_k (M)$ from only the knowledge
of the discrete sample of points $D \subset M$?
The strategy persistence uses to answer this question
is the following: Fix a distance parameter $\epsilon > 0$,
and build a simplicial complex $C_{\epsilon}$ by joining
a $m$-simplex whenever $m + 1$ data points in $D$ are mutually within
distance $\epsilon$ of each other. In this way, when
$\epsilon$ is sufficiently small (but not too small)
and $D \subset M$ is a sufficiently
dense sampling of the submanifold $M$, the complex $C_{\epsilon}$
is guaranteed to have the
same homotopy type (and thus the same sequence of homology groups)
as $M$. For a given fixed data set $D$, however, if $\epsilon$ is chosen
smaller than a threshold value
$C_{\epsilon}$ becomes a discrete set (with uninteresting topology), and,
similarly, if $\epsilon$ is larger than some other threshold value $C_{\epsilon}$
becomes a single giant simplex whose topology gives no information about $M$.
The true topology of $M$ is reflected in the simplex $C_{\epsilon}$ only when
$\epsilon$ ranges in an ``optimal"
interval between these two thresholds. The idea of persistence
is to inspect the variation in the topology of $C_{\epsilon}$ as $\epsilon$ varies,
and identify the largest interval in which the topology is ``persistent"
as a function of $\epsilon$. This persistent
topology is then the statistical estimate of
the ``true" topology of the data $D$.

While persistence relies on sophisticated constructions derived from algebraic 
topology, Morse Theory supplies the set of tools for an alternative
approach to topological data analysis~\cite{bremer,cazals,connolly}. The latter
provides a framework conducive to statistical analysis, because a probabilistic
estimate of the topology follows naturally.  
There exists a large body of knowledge on the applications of 
Morse Theory~\cite{nicol}. Although some of these works are quite complex and
sophisticated, to the knowledge of the present authors, some of the simpler
applications have not appeared in data analysis circles. In this paper,
a few synthetic examples of 3-dimensional point cloud data are 
utilized to illustrate these simple applications of Morse Theory. 

Morse Theory, in its simplest form, can be thought of as a set of topological 
constraints which must be satisfied by a surface, if/when some function on the 
surface is known. For example, consider a circle (i.e., a 1-dimensional,
compact surface) in 3 dimensions, oriented along the conventional z-axis.
Also, consider the height function on such a circle; it is a function defined
on the circle which produces the height of every point on the circle from the 
x-y plain. Such a function has two critical points, at the bottom and at
the top of the circle, where its derivative is zero. As shown in the
next section, these critical points of the height function restrict
the topology of the surface over which the function is defined. More
specifically, it is shown here that by computing the height function
and its critical points for point cloud data, Morse Theory allows one
to infer the genus of the underlying surface.  Moreover, 
resampling~\cite{efron,good}
is employed to compute the empirical sampling distribution of the genus,
which in turn allows for a probabilistic assessment of topology.

The contributions of this work are twofold. First, it is shown that Morse Theory
can be employed to infer the topology of the manifold underlying point cloud data.
In the examples, which are point clouds in $R^3$ , the manifolds are 2-dimensional
surfaces and their topology is uniquely set by one integer: the genus. 
Second, we point out that the genus (and more generally, all algebraic topological 
invariants of the data) must be treated as a random variable when inferred from 
data. A resampling method is employed to compute the empirical sampling 
distribution of the genus, which in turn, conveys its sampling variability. 
As such, one can predict the underlying topology in a probabilistic fashion.

\section{Method}

\subsection{Generalities}

To demonstrate the methodology, four compact surfaces are
considered (Figure 1). The choice of these specific examples is based on
the desire to have nontrivial, realistic, topology, but also sufficiently simple
topology to allow for a lucid presentation.  
The top/right panel in Figure 1 is topologically a sphere. However,
two ``dimples" are introduced in order to generate more critical points
for the height function, rendering the problem less trivial. The top/left panel shows the next nontrivial
example, namely a torus. These two surfaces have genus 0 and 1, respectively.
The next example (lower/left panel) is a genus 1 surface, but with 
``dimples," again for the purpose of having a more complex height function.
The final example (bottom/left) is a 2-torus, i.e., a genus 2 surface. Recall
that the goal is to infer these values of genus from data.

The particular embeddings/shapes
of the surfaces shown in Figure 1 are employed in the remainder
of the article. Other embeddings/orientations lead to different height
functions; alternatively, functions other than the height coordinate
can be used to assess the topology. The discussion section addresses the 
effect of changing the embedding
for the purpose of obtaining more precise (less variable) estimates of the genus.

Point cloud data are simulated by adding a zero-mean random Gaussian variable 
to the height function of the four surfaces. The variance of this variable
controls the level of noise in the data. Naturally, small values generally
lead to accurate and precise estimates of genus. Said differently, the 
inferred value of genus is the correct one, and the uncertainty of the
estimate is small. Although larger values of the variance are
associated with less precise estimates of genus, for sufficiently large
values the estimates become inaccurate as well, in the sense that the most likely
genus inferred from data is the wrong genus altogether. An analysis of the
sensitivity of the method to noise level is sufficiently complex to be
relegated to a separate article (reported later). The complexity of that 
analysis arises
because the effect of noise level is confounded with the relative size of 
the various loops around the handles.
For example, even with low noise levels, if one of the tori
in the 2-torus is much smaller than the other, then the method is likely
to imply that the underlying surface has genus one. 
For the present work, suffice it to say that the standard deviation of the
noise is fixed at 0.1. Loosely speaking, given that the radius of the
small loop in the torus example is 4 (grid lengths), a standard deviation of 
0.1 corresponds to a signal to noise ratio of about 40.

As shown in the next section, Morse Theory can place bounds on the
genus of a surface from knowledge of the critical points of a function
defined on the surface. Specifically, what is required is the number of 
minima, saddle points, and maxima. There exist numerous standard methods
for finding critical points of a function, but in this article a relatively
simple approach is adopted for the sake of clarity. 

\subsection{Specifics}

Although the height function is a standard function on a 
surface~\cite{bott1,nicol},
the function adopted in this article is the area of the surface
up to some height $h$, denoted $S(h)$. The area function is closely 
related to the height function, but is more natural when dealing with 
data. First, the height function for data is more noisy than the area 
function, because the latter is inherently an integral. Second,
critical points of the height function correspond to points in $S(h)$
where the derivative $S'(h)$ is discontinuous. The more robust nature,
and the presence of ``kinks" in the area function make it a natural 
choice to use for identifying the critical points in the height function.

Given that $S(h)$ is computed from data, it is a random variable. In 
other words, every realization of the Gaussian about the surface will
lead to a different value. In order to assess the variability of $S(h)$
resampling is employed~\cite{efron,good}. Specifically, 
100 samples/realizations are 
drawn and the distribution of $S(h)$, at prespecified values of $h$ is 
generated. Each distribution is summarized with a boxplot and displayed
for all $h$ values as a means of displaying the functional dependence
of $S(h)$, as well as its variability, on $h$. 

Note that each sample/realization of data gives rise to a sequence of
$S(h)$ values at prespecified $h$ values. As such, $S(h)$ can be considered
a stochastic 
time series. Additionally, it is a monotonic, non-decreasing time series.
This monotonic nature of the time series makes it difficult to identify
its kinks (i.e., critical points of the height function). A more useful 
quantity is the first derivative of
$S(h)$ with respect to $h$. Second derivatives are also useful, but
here only the time series of the first derivatives, $S'(h)$, is examined.  
It is the critical points of the $S'(h)$ time series which are used
in Morse Theory to infer topology. The sampling variability of $S'(h)$ is 
again assessed via resampling, and displayed with boxplots. 

Figure 2 shows the above ideas for the specific example of
a dimpled sphere. The top/left panel shows a vertical cross-section
of the surface. Here the $h$ values vary from the global minimum of the 
surface to its global maximum, in increments of 0.5.
The data simulated about this surface are not shown,
but boxplots summarizing the distribution of the $S(h)$ are shown
in the top/right panel. Although the boxplots are relatively small,
and difficult to see, their medians are connected by a red line
as a visual aid. Also difficult to see are the ``kinks" in the
red line at the critical points, marked
by the blue horizontal lines. The first derivative (left/lower panel)
better shows both the kinks and the sampling variability. It is
evident that some kind of a kink exists at each of the critical points of $S'(h)$
(again, marked by the blue lines). The kinks can be
generally classified into three types: an increasing step function, 
a cusp (i.e., $\wedge$), and a decreasing step function, respectively 
corresponding to 
minima, saddle points, and maxima. The second derivative of $S(h)$
is also shown (lower/right panel), only to illustrate that it too
carries information useful for identifying critical points. However, 
it is not used in the present work.  

The analogous figures for the torus example are shown in Figure 3.
Again, it can be seen that the kinks in the area function (and its
derivatives) occur at the locations of the critical points of $S'(h)$,
and that the shape of the kinks in the first
derivative are of the same type as seen previously, namely step
functions, and cusps. Similar results are found for the dimpled
torus and the 2-torus (not shown).

\subsection{Finding Critical Points}

Although there exist standard methods for finding critical points
of a time series, most rely on some sort of time series modeling. 
The time series models, in turn, have numerous parameters which must
be determined. Although there exist criteria (e.g., maximum likelihood)
for estimating the best models, for the sake of clarity,
a very simple approach is adopted here. The approach is based on
template matching. Specifically, three templates are selected
corresponding to the aforementioned three kinks observed in the series $S'(h)$,
namely 1) an increasing step function for finding local minima in the time
series, 2)  a cusp function for finding the saddle points, and 3) a decreasing 
step function for identifying local maxima in the series.

By sliding each of the templates across the time series for $S'(h)$,
and computing the residuals, one obtains three additional time series.  
The left column in Figure 4 shows these series for one realization of 
data about the dimpled sphere. The vertical lines
are at the $h$ values corresponding to the critical points. Given that
these time series are of residuals, near-zero values indicate a close
agreement between the template and the time series of $S'(h)$. It can
be seen that the residuals corresponding to the first template (top/left
panel in Figure 4) approach zero only at the location of the local minima.
Similarly, the residuals for the second template (middle/left panel)
are near zero only at the location of the saddle points. The final
panel shows the residuals for the last template, and the
residuals are near zero only at the location of the local maxima. 
To quantify the notion of ``near-zero," the histogram
of the residuals is examined (right column in Figure 4). Specifically, 
any residual within one standard deviation of zero is defined to be
``near-zero." This 1-standard-deviation value is displayed with the
vertical line on the histograms in Figure 4.

In short, sliding three templates across the time series of $S'(h)$, and
examining near-zero values of the ensuing residuals correctly identifies 
the locations of the critical points of $S(h)$. This method
for automatically identifying critical points of the height function
for data can be improved upon.  And as mentioned previously, there exist
more sophisticated methods for identifying critical points. However, that
is not the main goal of the present work. The rudimentary method
outlined here is sufficient to demonstrate the main goal of the work - 
that Morse Theory can be employed to estimate the topology underlying data,
and to express the statistical uncertainty in that estimate.

\section{Morse Theory}

The material presented in this 
subsection is only a small portion of Morse Theory, and so, has
been called Baby Morse Theory~\cite{bott1,bott2}.  

Given a surface $S$, the Poincare polynomial is defined as
\[ P(S) = \sum_k \; b_k \; t^k \;,\]
where  $-1 \leq t  \leq 1$, is a quantity with no special meaning,
and $b_k$ is the $k^{th}$ Betti number. For a 2-dimensional surface, $k = 0, 1, 2$.
Intuitively, $b_0$ is the number of simply-connected components of $S$, 
$b_1$ is the number of noncontractable loops on the surface, and $b_2$ is
the number of noncontractable surfaces. For example, for a 2-sphere,
$P(S) = 1 + t^2$, and for a torus, $P(S) = 1 + 2t + t^2.$ The $2t$ term
reflects the fact that there are two noncontractable loops on a torus - one
around the ``hole" of the doughnut, and another going around the ``handle." As
another example, consider a 2-torus for which $P(S) = 1 + 4t + t^2$. It is
important to point out that $P(S)$ is a topological quantity in the
sense that any 2-sphere (symmetric, squashed, dimpled, or otherwise)
has $P(S) = 1 + t^2$. The same is true of the other examples considered; 
their Poincare polynomial is independent of their embedding/shape.

Given a function $f$ defined on a surface, the Morse polynomial is
defined as
\[ M(f) = \sum_{P_i} \; t^{n_i} \;\;,\]
$P_i$ denotes the critical points of $f$, and $n_i$ is the index of $f$ at 
the $i^{th}$ critical point. The index is defined to be the number of non-decreasing
directions for $f$. Unlike the Poincare polynomial, the Morse function is
not a topological quantity. For example, consider the perfectly round 2-sphere.
Then the height function has 2 critical points, with indices
0 and 2, corresponding to the South and North poles, respectively. This is so,
because at the South pole there are no directions in which the height
function decreases, while there are two such directions at the North pole.
Then, for the height function on this sphere one has $M(f) = 1 + t^2$.
By contrast, a 2-sphere with dimples in it (e.g., top/left panel in Figure 1) 
has 6 critical points with indices 0, 1, 2, 0, 1, 2, respectively,
moving up from the bottom of the figure. For this height function,
$M(f) =2 + 2t + 2t^2.$  As another example, for the height function on 
the torus in the top/right panel of Figure 1, one has $M(f) = 1 + 2t + t^2$.

Central to Morse Theory are the so-called Morse inequalities~\cite{bott1,nicol}.
They are expressed in two forms - ``weak" and ``strong:"
\be M(f) \geq P(S)  \;\;\;\;, \;\;\;\; M(f) - P(S) = (1+t) Q(t), \ee
where $Q(t)$ is any polynomial in $t$ with non-negative coefficients.

In the above examples, note that for some functions one has $M(f) = P(S)$. 
Such functions are called ``perfect." Intuitively, such a function tightly 
``hugs" the surface. As such, the coefficients in the corresponding
Morse function are equal to the Betti numbers. As a result, knowledge
of a perfect function is tantamount to precise knowledge of the topology
(technically, homology)
of the underlying surface. For all non-perfect functions, the Morse
inequalities provide only an upper bound on the Betti numbers, and do
not uniquely identify the topology.

The search for perfect functions is aided by the Lacunary principle~\cite{bott1}:
If the product of all consecutive coefficients in $M(f)$ is zero, then $f$ is 
perfect. Another useful corollary of the strong form of the inequalities
follows upon considering $t=-1$: 
\be \sum_{P_i} (-1)^{n_i}  = \sum_k \; b_k \; (-1)^{k}. \ee
This places a constraint on the allowed number of minima, saddle points, and maxima:
\be n_{min} - n_{saddle} + n_{max} = b_0 - b_1 + b_2. \ee
And since in this article only surfaces with $b_0=b_2=1$ are considered, then
\be b_1 = 2 - n_{min} + n_{saddle} - n_{max}\;.\ee
Finally, given that any 
2-dimensional surface can be constructed by gluing tori to a sphere, it
follows that $b_1$ must be even (including zero). The genus of a compact
surface is then found to be 
\be \mbox{genus} = b_1/2\; . \ee
For a sphere, a torus, and a
2-torus (e.g. in Figure 1), the genus is 0, 1, and 2, respectively. Intuitively,
the genus counts the number of ``holes" or ``handles" in a compact surface.  
Inferring the genus is the main goal of the present work.

\section{Results}

Armed with a method to find the number of minima, saddle points, and maxima of 
the height function (section 2.3), one can then examine the distribution
of each. The top/left panel in Figure 5 shows the boxplots summarizing
the three distributions for the dimpled sphere example. Recall that
for this example, the correct number of minima, saddle points, and
maxima is 2. The median of the three boxplots is precisely at 2, as well.
The 1st and 3rd quartiles of the distribution (i.e., the bottom
and top sides of the boxes) suggest an uncertainty of about $\pm 1$
for each of the numbers. In other words, the number of minima, 
saddle points, and maxima generally varies within 1 of the correct
value (i.e., 2).

However, not all of the values in that range are allowed. Eq. (4)
constraints the three numbers, because the first Betti number must be even.
This constraint reduces the uncertainty even further. Meanwhile, the main
interest is in the value of the genus, which can be
computed from Eq. (5). The histogram of the genus
is shown immediately below the comparative boxplots in Figure 5.
Interpreting this histogram probabilistically, it can be seen that 
the most likely value of the genus is zero. And, of
course, that is the correct value. Moreover, values of estimated genus as large
as 2 are possible, but less likely.

The remaining panels in Figure 5 show the analogous figures
for the torus (top/right), dimpled torus (bottom/left), and
the 2-torus (bottom/right).  The correct number of minima,
saddle points, and maxima for the torus is (1,2,1). The analogous
numbers for the dimpled torus and the 2-torus are (3,6,3),
and (1,4,1), respectively. The comparative boxplots in Figure 5,
are all in agreement with these numbers. It is worth noting
that the width of the boxplots generally increases with the
complexity of the underlying surface. 

The distribution of the estimated genus for all four examples
is also consistent with the correct values. The most likely
genus for the torus, dimpled torus, and 2-torus are 1, 1, and
2, respectively - the correct values. As with the number of critical points,
the uncertainty in the estimated genus increases
with the complexity of the surface. Whereas the genus
for the dimpled sphere varies between 0 and 2, the range
for the 2-torus is 0 to 6. 

\section{Summary and Discussion}

The Morse inequalities are reviewed. It is shown that when specialized to the 
case of a 2-dimensional surface embedded in 3 dimensions, they place severe 
constraints on the topology of the surface. Three examples are employed to 
show that all of the quantities appearing in the Morse inequalities can be 
estimated from point cloud data, thereby providing a statistical/probabilistic
view of the topology of the surface underlying the data. Empirical sampling
distributions are produced for the various topological entities, all of which
can then lead to traditional confidence intervals or hypothesis tests of
the topological parameter of interest.  Throughout the
paper, an attempt is made to avoid complex mathematics (e.g., algebraic
and differential topology, and homology), with the hope that the utility
of Morse Theory in data analysis may be appreciated by a wider readership.

As mentioned above, the sensitivity of the inferred quantities to
noise level has not been examined here. The main reason is that 
the noise level and the physical size of the structures underlying
the point cloud data are confounded. This complication is not
unsurmountable; it simply calls for a more careful analysis wherein
the size of the noise and the typical features in the data are
both varied/controlled. 

The typical size of the features in the data also affects the
uncertainty of the inferred topology. The empirical sampling distribution
of the genus spreads out when the topological features are small
relative to noise level. Although not shown here, we have found that
this uncertainty depends on the orientation of the surface. This is
expected, because the height function depends on the orientation.
So, it is possible to orient the surface in a way that would allow
for more precise estimation of the critical points. In other words, 
it is possible to add another step to the proposed method, wherein
the variance of the distribution of genus is minimized across different
orientations of the point cloud. Such a rotation can also be used to
identify a perfect height function, in which case the Betti numbers
can be computed precisely, as opposed to being only bounded at the top.
This idea will be examined at a later time.

In the examples considered here the goal is to identify the
genus of a 2-dimensional compact surface underlying 3-dimensional
point cloud data. Several generalization are possible. The dimensionality
of the embedding space, or of the ``surface" (embodying the underlying
structure), can both be generalized. Of course, a single number like
genus will no longer suffice to define the topology uniquely, but the
set of Betti numbers does. In other words, if the manifold of interest
underlying the data has dimension larger than 2, then more parameters
need to be estimated. From a statistical point of view, the consequence of 
this increase in the number of parameters is that more data will be
required to estimate the parameters with precision.

The general formulation of Morse Theory does not require the underlying 
manifold to be compact. There are also extensions of Morse Theory that
allow for degenerate critical points, as well as extensions to manifolds with
boundary, and to Morse functions that take values in more general
spaces than $\bf R$ (e.g., circle-valued Morse Theory where Morse functions
are $S^1$-valued)~\cite{pajitnov}. The application of these more powerful
topological tools to data analysis is a fruitful frontier for exploration.


\textbf{Acknowledgements}

Marina Meil\v{a} is acknowledged for valuable discussion.
support for this project was provided by the
National Geospatial-Intelligence Agency, award number HM1582-06-1-2035.


\section{References}

\newpage
$ $
 
\vspace{-1.0in}


\centerline{ \includegraphics[height=9.5in,width=8in]{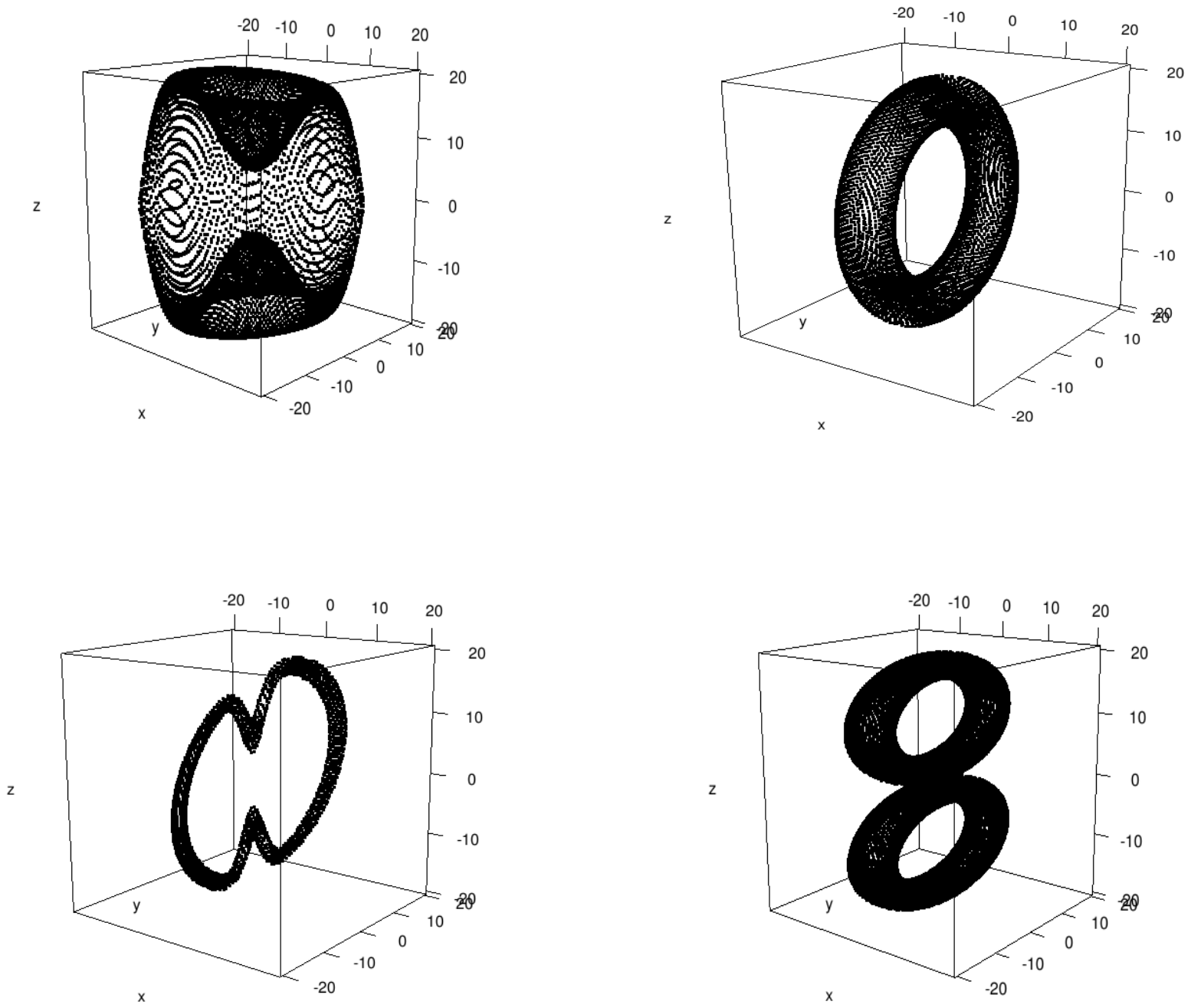} }
           
\vspace{-3.0in}
 
Figure 1. Four example surfaces: The ``dimpled sphere" in the top/left
panel is topologically a sphere (genus = 0), in spite of the dimples.
The top/right panel shows a surface with less trivial topology, namely
a torus (genus = 1). Another example of a genus-1 surface is shown
in the lower/left panel; it is a skinny torus, with ``dimples" at the
top and the bottom. The last example is a 2-torus (genus = 2) shown
in the lower/right panel.

\newpage

\centerline{ \includegraphics[height=6in,width=6in]{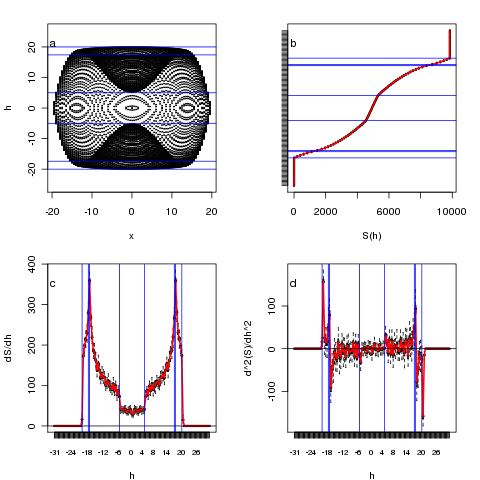}}

Figure 2. a) A vertical cross-section of the dimpled sphere shown in Figure 1. 
The blue lines mark the height of the critical points.
b) The dependence of the area function $S(h)$ on the height $h$ shown along
the y axis. The blue horizontal lines mark the height of the critical points.
c) The first derivative of $S(h)$ with respect to $h$, and the second
derivative in panel d).

\newpage

\centerline{ \includegraphics[height=6in,width=6in]{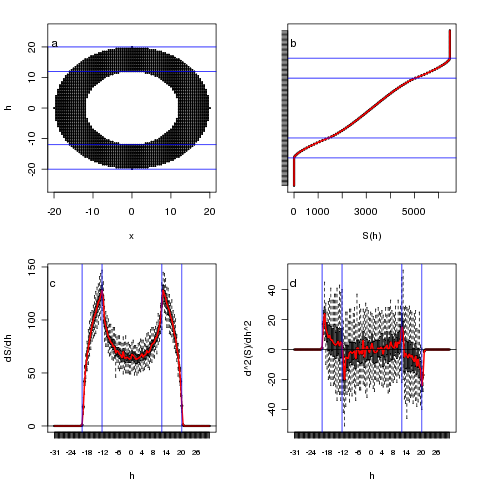}}

Figure 3. Same as Figure 2, but for the torus.

\newpage

\centerline{ \includegraphics[height=6in,width=6in]{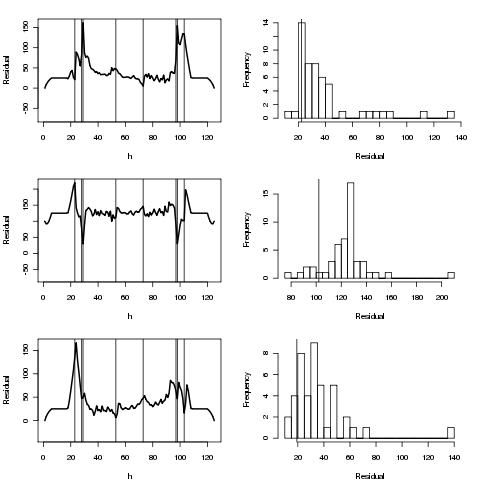}}

Figure 4. Left column: The time series generated by sliding three
template across the time series of $S'(h)$ and computing a
measure of the error/residual between the time series and each
template. Right column: The histogram of the three template
errors. From top to bottom, the templates are the increasing
setp function, the cusp, and the decreasing step function. 

%
%

%
%
%
\newpage
$ $

\vspace{-1.0in}

%
%

\centerline{ \includegraphics[height=9.5in,width=8in]{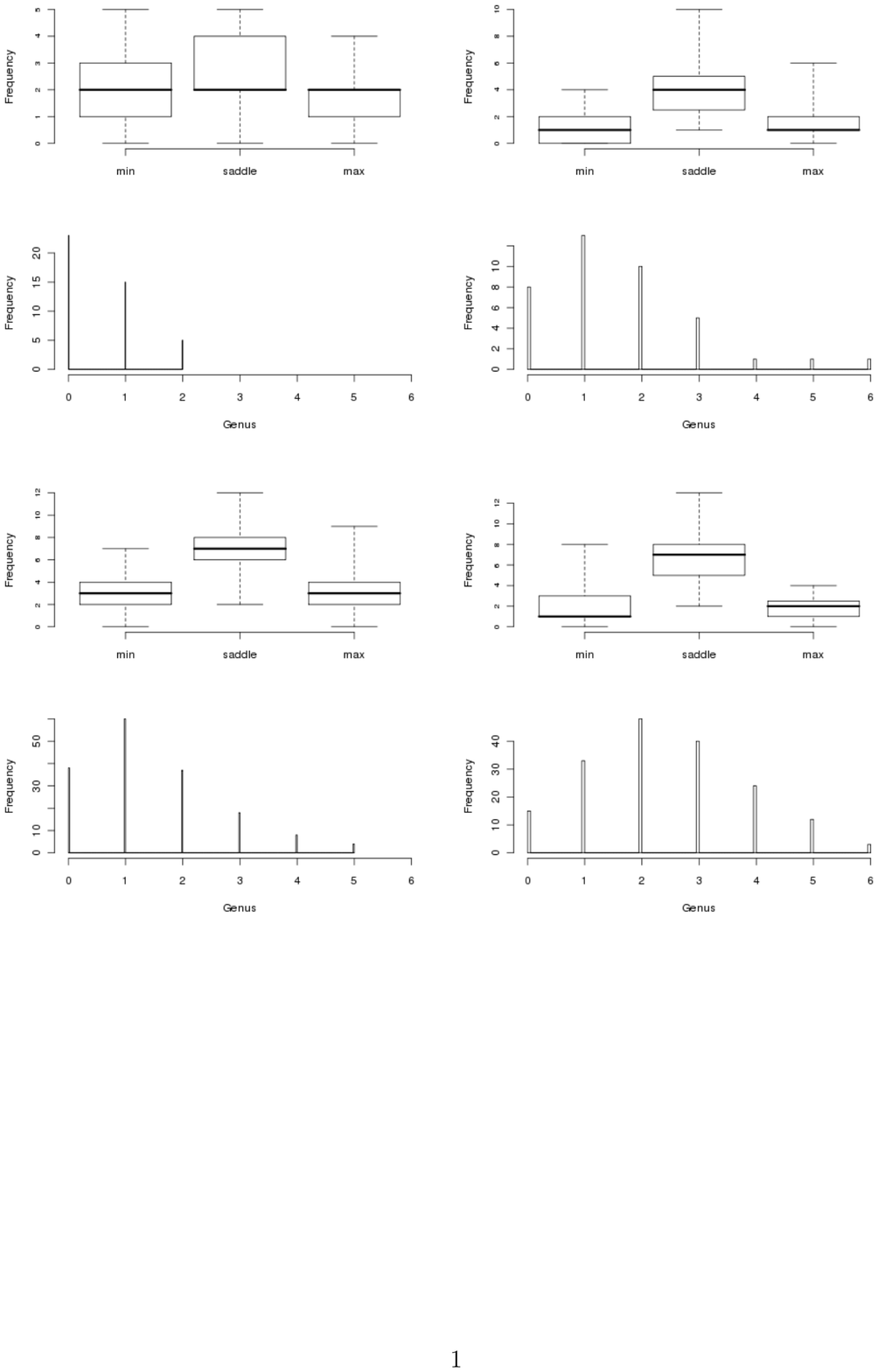} }

\vspace{-3.0in}
 
Figure 5.  The boxplot summary of the distribution of the number of minima, 
saddle points, and maxima, and the histogram of the estimated genus.
The 4 panels pertain to the four examples: dimpled sphere (top/left),
torus (top/right), dimpled torus (bottom/left), 2-torus (bottom/right).


\begin{thebibliography}{99}

 \bibitem{adan} A. Adan, C. Cerrada, V. Feliu, Modeling Wave Set: Definition and 
Application of a New Topological Organization for 3D Object Modeling,
Computer Vision and Image Understanding, 79 (2000) 281-307.

 \bibitem{bott1} R. Bott, Morse theoretic aspects of Yang-Mills Theory, in
Recent Developments in Gauge Theories, Eds. G. 'tHooft, et al., Plenum Publishing, 
1980, pp. 46-67.

 \bibitem{bott2} R. Bott,  Lectures on Morse Theory, Old and New, 
Bull. Amer. Math. Soc. (N.S.), 7 (1982) 331-358.

 \bibitem{bremer} P-T.\ Bremer, V. Pascucci, A Practical Approach to Two-Dimensional
Scalar Topology, in Topology-based Methods in Visualization,
H.\ Hauser, H.\ Hagen, H.\ Theisel (Eds), Springer-Verlag, Berlin, 2007.
ISBN-13 978-3-540-70822-3.

 \bibitem{carlsson} G. Carlsson, Topology and Data, Bull. (New Series) Amer. Math.
Soc., 46 (2009) 255­308.

\bibitem{cazals} F. Cazals, F. Chazal, T. Lewiner, Molecular Shape Analysis 
Based upon the Morse-Smale Complex and the Connolly Function.  Proc. 19th
ACM Symp. Computational Geometry (SoCG), 2003, pp. 351-360.

\bibitem{connolly} M. Connolly, Molecular Surfaces: A Review, Network Science, (1996) .

 \bibitem{efron} B. Efron, R. J. Tibshirani, An introduction to the
bootstrap, Chapman \& Hall, London, 1993.

 \bibitem{everitt} B. S. Everitt, Cluster Analysis, Second Edition,
Heinemann Educational Books. London, 1980.

 \bibitem{gilmore} R. Gilmore, M. Lefranc, The Topology of Chaos,
Wiley-Interscience, New York, 2002. ISBN 0-47 1-40816-6

 \bibitem{good} P. I. Good, Introduction to Statistics Through Resampling
Methods and R/S-PLUS, Wiley, 2005. ISBN 0-471-71575-1

 \bibitem{grejner} D. Grejner-Brzezinska, C. Toth, Deriving Vehicle Topology and Attribute 
Information Over Transportation Corridors From LIDAR Data. Proceedings of the 59th 
Annual Meeting of The Institute of Navigation and CIGTF 22nd Guidance Test 
Symposium, June 23 - 25, Albuquerque, NM., 2003, pp. 404-410.

 \bibitem{haji} M. R. Hajihashemi, M. El-Shenawee, Shape Reconstruction Using the 
Level Set Method for Microwave Applications, Antennas and Wireless Propagation 
Letters, 7 (2008) 92-96.

 \bibitem{katz} R. W. Katz, A.H. Murphy, Economic Value of Weather and Climate
Forecasts, Cambridge University Press, Cambridge, 1997.

 \bibitem{lee} J.\ M. Lee, Introduction to Topological Manifolds,
Springer-Verlag, New York, 2000. ISBN 0-387-98759-2

 \bibitem{leever} J. A. Lee, M. Verleysen, Nonlinear dimensionality reduction,
Information Science and Statistics, Springer, 2007.

 \bibitem{nicol} L. I. Nicolaescu, An Invitation to Morse Theory,
Springer Monograph, XIV, 2010. ISBN 978-0-387-49509-5. 


 \bibitem{niyogi} P. Niyogi, S. Smale, S. Weinberger, Finding the homology of 
submanifolds with high confidence from random samples, Combinatorial and 
Discrete Geometry, 39 (2008) 419-441.

 \bibitem{pajitnov} A. V. Pajitnov, Circle-valued Morse Theory,
De Gruyter Studies in Mathematics 32, 2006. ISBN 978-3-11-015807-6.

 \bibitem{pasc} V. Pascucci, X. Tricoche, H. Hagen, J. Tierny, Topological
Methods in Data Analysis and Visualization; Theory, Algorithms, and Applications.
Springer, 2010. 

 \bibitem{waw} T. F. Wawrzyniec, L. D. McFadden, A. Ellwein, G. Meyer, L. Scuderi, 
J. McAuliffe, P. Fawcett, Chronotopographic analysis directly from 
point-cloud data: A method for detecting small, seasonal hillslope change, 
Black Mesa Escarpment, NE Arizona, Geosphere, 3, (2007) 550-567. 
DOI: 10.1130/GES00110.1

 \bibitem{zomo} A. Zomorodian, Topology for Computing, Cambridge Monographs on 
Applied and Computational Mathematics (No. 16), 2005.

\end{thebibliography}
\end{document}